\documentclass[preprint,12pt]{elsarticle}

\usepackage[T1]{fontenc}
\usepackage{booktabs}
\usepackage{amssymb}
\usepackage{amsmath}
\usepackage{bbm}
\usepackage{bm}
\usepackage{color}
\usepackage{hyperref}
\usepackage{comment}
\usepackage{courier}
\usepackage{natbib}
\journal{Computational Statistics and Data Analysis}

\providecommand{\normaldistn}{\operatorname{Normal}}
\providecommand{\unifdistn}{\operatorname{Unif}}
\providecommand{\stderr}{\widehat{\mathrm{se}}}
\providecommand{\seKR}{\stderr_\mathrm{KR}}
\providecommand{\ciN}{\mathrm{CI}_\mathrm{N}}
\providecommand{\ciS}{\mathrm{CI}_\mathrm{S}}
\providecommand{\ciKR}{\mathrm{CI}_\mathrm{KR}}
\providecommand{\ciPone}{\mathrm{CI}_{\mathrm{p}1}}
\providecommand{\ciPtwo}{\mathrm{CI}_{\mathrm{p}2}}
\providecommand{\dfPB}{\nu_\mathrm{PB}}
\providecommand{\dfS}{\nu_\mathrm{S}}
\providecommand{\dfKR}{\nu_\mathrm{KR}}

\begin{document}
\sloppy
\begin{frontmatter}



\title{A studentized permutation test for the treatment effect in individual participant data meta-analysis} 


\author[label]{Phuc Thien Tran} 
\author[label]{Long-Hao Xu} 
\author[label]{Christian R\"{o}ver} 
\author[label]{Tim Friede} 

\affiliation[label]{organization={Department of Medical Statistics, University Medical Center G\"{o}ttingen},
            addressline={Humboldtallee~32}, 
            city={G\"{o}ttingen},
            postcode={37073}, 
            country={Germany}}

\begin{abstract}

Meta-analysis is a well-established tool used to combine data from several independent studies, each of which usually compares the effect of an experimental treatment with a control group. While meta-analyses are often performed using aggregated study summaries, they may also be conducted using individual participant data (IPD). Classical meta-analysis models may be generalized to handle continuous IPD by formulating them within a linear mixed model framework. IPD meta-analyses are commonly based on a small number of studies. Technically, inference for the overall treatment effect can be performed using Student-t approximation. However, as some approaches may not adequately control the type I error, Satterthwaite’s or Kenward-Roger’s method have been suggested to set the degrees-of-freedom parameter. The latter also adjusts the standard error of the treatment effect estimator. Nevertheless, these methods may be conservative. Since permutation tests are known to control the type I error and offer robustness to violations of distributional assumptions, we propose a studentized permutation test for the treatment effect based on permutations of standardized residuals across studies in IPD meta-analysis. Also, we construct confidence intervals for the treatment effect based on this test. The first interval is derived from the percentiles of the permutation distribution. The second interval is obtained by searching values closest to the effect estimate that are just significantly different from the true effect. In a simulation study, we demonstrate satisfactory performance of the proposed methods, often producing shorter confidence intervals compared with competitors.
\end{abstract}





\end{frontmatter}



\section{Introduction}
\label{intro}

Meta-analysis is a well-established method to combine, summarize, and analyze research results from several independent studies addressing a similar question \cite{hedges_olkin85,higgins_green11,schmid_et_al20}. In medicine, meta-analysis can be used to evaluate the effect of a treatment given the summary from the treatment- and the control-group. Meta-analysis is commonly performed using aggregate data from published reports, but may also be based on \emph{individual participant data (IPD)}. When the patient-level data (on outcomes and possibly covariates) are continuous measurements, an IPD meta-analysis may usually be performed using linear mixed model methodologies \citep{RileyTierneyStewart}.  

A motivating example is given by the investigation by Kambi\v{c} and colleagues \cite{kambic_et_al24}, who recently conducted an IPD meta-analysis on the safety and efficacy of exercise training on exercise capacity (e.g., six-minute walk test distance) and quality of life compared to standard care in left ventricular assist device patients. 
The meta-analysis encompasses four studies, each of which included between~$14$ and~$54$ patients, and is performed by assuming a linear mixed model. Since the meta-analysis includes only a small number of studies with limited sample sizes, the significance test and the $95\%$ confidence interval for the treatment effect based on the asymptotic normal distribution of the effect's estimator might not control the type~I error adequately \cite{pinheiro00,legha_et_al18}. On the other hand, Satterthwaite's approach and Kenward-Roger's approach, which are well-known Student-$t$-based approaches in linear mixed models, might be too conservative \cite{legha_et_al18}. Since permutation tests control the type~I error well, even for small sample sizes, they might serve as a useful alternative also in IPD meta-analysis. 

To our knowledge, permutation tests have not been applied successfully in IPD meta-analysis. \citet{basso_finos12} propose a test for generalized linear mixed models. 
However, this test is equivalent to the permutation approach suggested by \citet{follmann_proschan04} for the treatment effect in classical meta-analysis, which is not recommended when the number of studies is small. 
If the assumed linear mixed model can be written in form of a generalized linear model (GLM), several permutation techniques exist for testing regression coefficients \cite{ogorman01,kheradpajouh_renaud10,winkler_et_al14,nyblom15}. However, these approaches cannot be used to test the treatment effect directly, since residuals in the rewritten model might not be exchangeable in general due to unequal residual variances. 



In this paper, we propose a studentized permutation test for IPD meta-analysis of continuous outcomes, even for small numbers of studies and the small study sizes. A crucial feature is that this test assumes the exchangeability of the \emph{standardized} residuals rather than the usual (``plain'') residuals.
Standardized residuals are permuted, and a $t$-statistic is computed based on these residuals. 
In addition, the proposed test allows for permutation across studies in an IPD meta-analysis, and it does not rely on the assumption of normality. Based on this test, a confidence interval for the treatment effect is constructed (using the quantiles of the permutation distribution), which is consistent with the proposed test. Motivated by \citet{follmann_proschan04}, an alternative confidence interval is computed by searching for values that are closest to the effect estimate and just significantly different from the true effect.

The paper is structured as follows. Section~\ref{sec:preliminary} presents a commonly used linear mixed model in IPD meta-analysis of continuous outcomes, along with standard approaches for constructing tests and confidence intervals for the treatment effect. Section~\ref{sec:method} shows the idea and detailed algorithm of the proposed studentized permutation test and two confidence intervals. Section~\ref{sec:sim} presents a simulation study where the proposed methods are evaluated and compared to existing methods under various scenarios. Section~\ref{sec:discussion} concludes the paper with a brief discussion.

\section{Statistical model and standard approaches}\label{sec:preliminary}
\subsection{Conventional and novel approaches}
In this section, we establish mathematical notation and introduce popular methods for estimation and inference within an IPD meta-analysis framework. The introduced model essentially falls into the class of linear mixed models (LMMs) for continuous outcomes. We discuss the advantages and disadvantages of these methods, providing the motivation for our subsequently proposed approaches.

\subsection{Parametric model for IPD meta-analysis}
\label{ipd-ma}

Consider a meta-analysis of $k$ studies. Let $y_{ij}$ denote the continuous outcome of patient $j$ in the $i$-th study ($j=1,\ldots,n_i, \ i=1,\ldots,k$), $y_{0ij}$ denote the baseline value of the outcome, and $z_{ij}$ denote the group indicator (1 for treatment, 0 for control). Assume the following stratified-intercept random-effects model for $y_{ij}$:
\begin{equation}\label{eqn:mod}
y_{ij}=\beta_{0i}+\beta_{1i}y_{0ij}+u_{i}z_{ij}+\varepsilon_{ij},\\  
u_{i}\sim \normaldistn\left(\theta,\tau^2\right), \ \varepsilon_{ij}\sim \normaldistn\left(0,\sigma^{2}_{i}\right),    
\end{equation}
where~$\beta_{0i}$ and~$\beta_{1i}$ denote the intercept and slope stratified by study, $u_i$~denotes the random slope, $\theta$~denotes the treatment effect, $\tau^2$~denotes the between-study heterogeneity, and $\sigma_i^2$ is the residual variance in study~$i$ \cite{goldstein_et_al00,riley_et_al13,debray_et_al15,legha_et_al18,riley_debray21}.  
In fact, $\sigma_{i}$~is also referred to as the \emph{unit information standard deviation (UISD)} of study~$i$, since it relates the contribution of each participant to the likelihood of this study \cite{roever_et_al21}. 
The model in equation~(\ref{eqn:mod}) is commonly applied in practice; in particular, it is a linear mixed model and may be rewritten in matrix form as follows:
\begin{equation}\label{eqn:mod_mat}
\bm{Y}=\bm{X}\bm{\beta}+\bm{Z}\bm{u}+\bm{\varepsilon}, \ \bm{u}\sim \normaldistn(\theta \bm{1}_k,\tau^2 \bm{I}_k),\  \bm{\varepsilon}\sim \normaldistn(\bm{0},\bm{R}),    
\end{equation}
where $\bm{u}=(u_1,\cdots,u_k)^\top$ is a vector of random effects, $\bm{Y}=(y_{11},\cdots,y_{1n_1},\cdots,y_{k1},\cdots,y_{kn_k})^\top$ is a vector of outcomes, $\bm{\varepsilon}=(\varepsilon_{11},\cdots,\varepsilon_{1n_1},\cdots,\varepsilon_{k1},\cdots,\varepsilon_{kn_k})^\top$ is a vector of residuals, $\bm{\beta}=(\beta_{01},\cdots,\beta_{0k},\beta_{11},\cdots,\beta_{1k})^\top$ is a vector of stratified intercepts and slopes, $\bm{R}=\text{diag}(\sigma^{2}_{1},\cdots,\sigma^{2}_{1},\cdots,\sigma^2_k,\cdots,\sigma^2_k)$ is a diagonal matrix of residual variances, and
\begin{equation}
    \bm{X}=\left(\begin{matrix}
        1 & 0 & \cdots & 0 & y_{011} & 0 & \cdots & 0\\
        \vdots & \vdots & & \vdots & \vdots & \vdots & & \vdots\\
        1 & 0 & & \vdots & y_{01n_1} & 0 & & \vdots\\
        0 & 1 & & \vdots & 0 & y_{021} & & \vdots\\
        \vdots & \vdots & & \vdots & \vdots & \vdots & & \vdots\\
        \vdots & 1 & & \vdots & \vdots & y_{02n_2} & & \vdots\\
        \vdots & 0 & & \vdots & \vdots & 0 & & \vdots\\
        \vdots & \vdots & & \vdots & \vdots & \vdots & & \vdots\\
        \vdots & \vdots & & 0 & \vdots & \vdots & & 0\\
        \vdots & \vdots & & 1 & \vdots & \vdots & & y_{0k1}\\
        \vdots & \vdots & & \vdots & \vdots & \vdots & & \vdots\\
        0 & 0 & \cdots & 1 & 0 & 0 & \cdots & y_{0kn_k}      
    \end{matrix}\right), \ \bm{Z}=\left(\begin{matrix}
        z_{011} & 0 & \cdots & 0\\
        \vdots & \vdots & & \vdots\\
        z_{01n_1} & 0 & & \vdots\\
        0 & z_{021} & & \vdots\\
        \vdots & \vdots & & \vdots\\
        \vdots & z_{02n_2} & & \vdots\\
        \vdots & 0 & & \vdots\\
        \vdots & \vdots & & \vdots\\
        \vdots & \vdots & & 0\\
        \vdots & \vdots & & z_{0k1}\\
        \vdots & \vdots & & \vdots\\
        0 & 0 & \cdots & z_{0kn_k}      
    \end{matrix}\right).\nonumber
\end{equation}
As a result, $\bm{Y}$~follows a multivariate normal distribution
\begin{equation}
  \bm{Y}\;\sim\;\normaldistn(\bm{X}\bm{\beta}+\theta \bm{Z}\bm{1}_k,\,\tau^2\bm{Z}\bm{Z}^\top+\bm{R})
\end{equation}
and the vector of parameters $(\theta,\beta,\tau^2,\sigma^2_1,\cdots,\sigma^2_k)^\top$ may be estimated using a maximum likelihood (ML) approach. The impact of estimating between-study heterogeneity on the treatment effect estimation can be mitigated by using the \emph{restricted maximum likelihood (REML)} method \cite{kackar_harville84}, which is a penalized estimation technique.
In the following, we consistently use REML method estimation throughout this paper.
We note that the REML estimators, denoted by $(\hat\theta,\hat\beta,\hat\tau^2,\hat\sigma^2_1,\cdots,\hat\sigma^2_k)^\top$, do not have any closed-form expressions and are computed via maximizing the restricted likelihood function numerically, e.g., using the Newton-Raphson method. 

Consider the following two-sided hypotheses:
\begin{equation}\label{eq:test}
H_{0}:\,\theta=\theta_0 \quad\mbox{vs.}\quad
H_{1}:\,\theta\neq \theta_0,
\end{equation}
where $\theta_0$ is known. When $\theta_0$ is non-zero, we simply need to subtract $\theta_0\bm{Z}\bm{1}_k$ from the data and test for a zero parameter as follows
\begin{equation}
H_{0}:\,\Tilde\theta=0 \quad \text{vs.} \quad
H_{1}:\,\Tilde\theta\ne 0,
\end{equation}
where $\Tilde{\theta}=\theta-\theta_0$. In this paper, unless stated explicitly, we assume $\theta_0=0$. Under the null hypothesis~$H_0$ and if $\theta_0=0$, the treatment has no effect, whereas under the alternative hypothesis~$H_1$, treatment and control groups differ.
This two-sided hypothesis test can be conducted using the test statistic
\begin{equation}\label{eqn:tstat}
    t=\frac{\hat\theta}{\stderr(\hat\theta)},
\end{equation}
where $\stderr(\hat\theta)$ denotes the estimated standard error of $\hat\theta$. Under regularity conditions, this statistic is asymptotically standard normal distributed. The corresponding $100\%(1-\alpha)$ Wald-type confidence interval $\ciN$ is given by $\hat\theta\pm z_{1-\alpha/2} \ \stderr(\hat\theta)$
where $z_{1-\alpha/2}$ denotes the $(1-\alpha/2)$-quantile of the standard normal distribution. The test based on the asymptotic distribution, however, does not necessarily the control type~I error and hence the confidence intervals tend to be too narrow \cite{pinheiro00,legha_et_al18}, in particular for small meta-analyses. 
Alternatively, Pinheiro and Bates \cite{pinheiro00} considered a Student-$t$ distribution with $\dfPB$~degrees of freedom as the reference distribution for testing the significance of fixed effects in linear mixed models. 
However, for the treatment effect in Model 
(\ref{eqn:mod_mat}), the degrees of freedom are $\dfPB=\sum_{i}n_i-2k$, which is large and hence yields a Student-$t$ approximation that is close to a standard normal distribution. 
Therefore, we will not consider Pinheiro's method further.

\subsection{Satterthwaite's approach}

The $t$-statistic in (\ref{eqn:tstat}) does not exactly follow a Student-$t$ distribution \cite{giesbrecht_burns85,kuznetsova_et_al17}. Based on Satterthwaite's method-of-moment approximation to the degrees of freedom \cite{satterthwaite46}, a Student-$t$ distribution that approximates the distribution of the $t$-statistic under the null hypothesis is suggested in \citet{giesbrecht_burns85}, 
where the degrees of freedom are 
computed based on the variance of the estimator of the treatment effect and the delta method.
The corresponding $100\%\left(1-\alpha\right)$ confidence interval $\ciS$ is given by $\hat\theta\pm t_{\dfS,(1-\alpha/2)}\,\stderr(\hat\theta)$,
where $t_{\dfS,(1-\alpha/2)}$ denotes the $(1-\alpha/2)$-quantile of the Student-$t$ distribution with $\dfS$~degrees of freedom. 

\subsection{Kenward and Roger's approach}

\citet{kackar_harville84} showed that the asymptotic covariance matrix of the estimated fixed effects $
(
\hat{\bm{\beta}},\hat{\theta})^
\top$ underestimates the true covariance matrix by a (matrix-valued) bias~$\Lambda$. Consequently, the standard error $\stderr(\hat{\theta})$ underestimates the standard deviation of $
\hat{\theta}$ and hence affects the inference on~$\theta$. Therefore, \citet{kenward_roger97} proposed an adjustment to the asymptotic covariance matrix by combining its bias-corrected modification with a Taylor expansion of the bias matrix~$\Lambda$. The adjusted standard error $\seKR(\hat{\theta})$ is obtained as the square root of the last diagonal element of the adjusted covariance matrix. Let $t_\mathrm{KR}$ denote the $t$-statistic resulting from this adjustment, \citet{kenward_roger97} approximate the distribution of the squared statistic~$t^{2}_\mathrm{KR}$ with an $F$-distribution with $1$~and $\dfKR$~degrees of freedom ($F_{1,\dfKR}$), where the degrees of freedom~$\dfKR$ are calculated by matching expectation and variance of the squared statistic~$t^{2}_\mathrm{KR}$ to the ones from~$F_{1,\dfKR}$. 
In testing the significance of the treatment effect~$\theta$, \citet{kenward_roger97} and \citet{halekoh_højsgaard14} show that the degrees of freedom in Kenward and Roger's approach are the same as those in Satterthwaite’s approach ($\dfKR = \dfS$). The corresponding $100\%(1-\alpha)$ confidence interval~$\ciKR$ is $\hat\theta\pm t_{\dfKR, (1-\alpha/2)}\,\seKR(\hat\theta)$,
where $t_{\dfKR,(1-\alpha/2)}$ denotes the $(1-\alpha/2)$-quantile of the Student-$t$ distribution with $\dfKR$~degrees of freedom. 

\subsection{Remark}

Satterthwaite's approach and Kenward-Roger's approach are commonly used for inference in linear mixed models, even for small sample sizes \cite{giesbrecht_burns85,kenward_roger97,halekoh_højsgaard14,kuznetsova_et_al17}. 
Although confidence intervals from Satterthwaite's approach and Kenward-Roger's approach exhibit similar empirical coverage probability close to the nominal level, they might be conservative \cite{legha_et_al18}. 
Moreover, by their formulation, Satterthwaite’s and Kenward-Roger’s approaches assume that $\bm{Y}$ in Model (\ref{eqn:mod_mat}) is normally distributed, which might be violated in practice.

\section{Proposed studentized permutation test and confidence interval}
\label{sec:method}

In the following, we first introduce a novel studentized permutation test. Then we discuss two different methods for constructing confidence intervals based on this test.
Separating fixed and random effects in the original linear mixed model~(\ref{eqn:mod_mat}) introduced above, it can be rewritten as follows:
\begin{equation}\label{eqn:mod_full}
\bm{Y}=\bm{X}\bm{\beta}+\theta\bm{Z}\bm{1}_k+\bm{\varepsilon}_0,\quad \bm{\varepsilon}_0\sim \normaldistn(\bm{0},\bm{\Sigma}_{\bm{\varepsilon}_0}),   
\end{equation}
where $\bm{\Sigma}_{\bm{\varepsilon}_0}=\tau^2\bm{Z}\bm{Z}^\top+\bm{R}$, $\bm{\varepsilon}_0=\bm{Z}\bm{u}_0+\bm{\varepsilon}$ is the residual, and $\bm{u}_0$ is the centered version of $\bm{u}$. Under the null hypothesis ($H_{0}:\theta=0$) the model simplifies to
\begin{equation}\label{eqn:mod_redu}
\bm{Y}=\bm{X}\bm{\beta}+\bm{\varepsilon}_0,\quad \bm{\varepsilon}_0\sim \normaldistn(\bm{0},\bm{\Sigma}_{\bm{\varepsilon}_0}).   
\end{equation}

As mentioned previously, the existing methods cannot be applied directly to~(\ref{eqn:mod_full}), since the elements of~$\bm{\varepsilon}_0$ in general are not exchangeable.
Inspired by \citet{lee_braun11}, we propose a method to address this problem; this approach aims to ensure exchangeability by using a structured transformation. Let~$\bm{W}$ denote the Cholesky decomposition of $\bm{\Sigma}_{\bm{\varepsilon}_0}$, the covariance matrix of~$\bm{\varepsilon}_0$. Then $\bm{W}$~is an upper triangle matrix such that $\bm{\Sigma}_{\bm{\varepsilon}_0}=\bm{W}^\top \bm{W}$. Let~$\bm{\Tilde{\varepsilon}}=(\bm{W}^\top)^{-1}\bm{\varepsilon}_0$ denote the standardized residuals in model~(\ref{eqn:mod_redu}). As a result, the covariance matrix of $\bm{\Tilde{\varepsilon}}$ is an identity matrix, and its elements~$\Tilde{\varepsilon}_{ij}$ are independent and identically distributed random variables. Therefore, under the null hypothesis, the exchangeability assumption of these residuals is satisfied, allowing them to be permuted across studies. 
The resulting algorithm for our proposed studentized permutation test is presented in Table~\ref{tab:algorithm}. A permutation $100\%(1-\alpha)$ confidence interval $\ciPone$ for $\theta$ can be constructed as follows:
\begin{equation}\label{ci-p}
    [\hat\theta-t_{1-\alpha/2}^\star\,\stderr(\hat\theta); \;
    \hat\theta-t_{\alpha/2}^\star\,\stderr(\hat\theta)],
\end{equation}
where $t_{(\alpha/2)}^\star$ and $t_{(1-\alpha/2)}^\star$ denote the permutation distribution's empirical $(\alpha/2)$- and $(1-\alpha/2)$-quantiles, respectively. 

\begin{table}[]
    \centering
    \caption{Pseudo code for the proposed studentized permutation test, generating a permutation distribution of $t$-statistic values.}
    \small
    \begin{tabular}{ll}
        \toprule
        1. & \texttt{Fit Model (\ref{eqn:mod_full}) to the response $\bm{Y}$ and compute }\\                                                              & $t$\texttt{-statistic}  $t=\hat{\theta}/\stderr(\hat{\theta})$.\\
        2. & \texttt{Fit Model (\ref{eqn:mod_redu}) to the response $\bm{Y}$ and compute}\\
           & \texttt{the standardized error} $\bm{\hat{\Tilde{\varepsilon}}}=(\bm{\hat W}^\top)^{-1}(\bm{Y}-\bm{X}\hat{\bm{\beta}}),$ \texttt{where $\bm{\hat W}$ is }\\
           & \texttt{the Cholesky decomposition of} $\bm{\hat{\Sigma}}_{\bm{\varepsilon}_0}=\hat{\tau}^2\bm{Z}\bm{Z}^\top+\bm{\hat{R}}$.\\
        3. & \texttt{For permutation $p=1,\cdots,N$,}\\
           & \hspace{0.5cm} - \texttt{Permute $\bm{\hat{\Tilde{\varepsilon}}}$ as $\bm{\hat{\Tilde{\varepsilon}}}_{(p)}$ and compute the permuted response}\\
           & \hspace{0.7cm}   $\bm{Y}_{(p)}=\bm{X}\bm{\hat{\beta}}+\bm{\hat W}^\top\bm{\hat{\Tilde{\varepsilon}}}_{(p)}$.\\
           & \hspace{0.5cm} - \texttt{Refit Model (\ref{eqn:mod_full}) to $\bm{Y}_{(p)}$ and compute $t$-statistic}\\
           & \hspace{0.7cm} $t_{(p)}=\hat{\theta}_{\left(p\right)}/\stderr(\hat{\theta}_{(p)})$.\\
             \bottomrule
    \end{tabular}
    \label{tab:algorithm}
\end{table}


This confidence interval (\ref{ci-p}) might have inflated type~I error when the treatment effect deviates substantially from zero. In this case, besides having different variances, residuals $\bm{\varepsilon}_0$ also have different means, so the standardized residuals $\bm{{\Tilde{\varepsilon}}}$ might not have a mean vector with constant components. As a result, the exchangeability condition is not satisfied and hence it is not valid to permute the standardized residuals. 
This is not a problem for the purpose of the test of the point hypothesis~$\theta=0$ (where exchangeability holds), but for constructing a confidence interval spanning a wider range of $\theta$~values, these should be considered explicitly in the calculation.

As a solution, inspired by the exact permutation-based confidence interval method in \citet{follmann_proschan04}, we propose a new method for constructing confidence intervals for treatment effect. This method is a search algorithm that finds values closest to the treatment effect estimate such that they are just significantly different from the true treatment effect, i.e., all values inside the resulted interval are insignificantly different from the true effect. This can be done by a search algorithm that finds values of boundary such that the hypothesis test which compares these values to the treatment effect rejects the null hypothesis of equality. It systematically adjusts the boundary value and performs the hypothesis test, ensuring that the final result satisfies the desired significance level.

Specifically, for the upper bound of the confidence interval, consider tests with a significance level $\alpha/2$:
\begin{equation}\label{eq:test2}
H_{0}:\,\theta=\theta_0 \quad \text{vs.} \quad
H_{1}:\,\theta<\theta_0
\end{equation}
Then we simply need to subtract $\theta_0\bm{Z}\bm{1}_k$ from the data and again test for a zero parameter (as in Equation \ref{eq:test}). Therefore, these tests are equivalent to testing
\begin{equation}
H_{0}:\,\Tilde\theta=0 \quad \text{vs.} \quad
H_{1}:\,\Tilde\theta< 0,
\end{equation}
where $\Tilde{\theta}=\theta-\theta_0$ and hence the procedure in Table \ref{tab:algorithm} can be applied. Starting from a value which is larger than $\hat\theta$, say $\hat\theta+z_{1-\alpha/2}\,\stderr(\hat\theta)$, $\theta_0$ is gradually increased until the test in (\ref{eq:test2}) rejects the null hypothesis, i.e., the resulting $p$-value does not exceed $\alpha/2$. In practice, a range for $\theta_0$ is set at the beginning, e.g., $[\hat\theta+z_{1-\alpha/2}\,\stderr(\hat\theta);\hat\theta+t_{2,1-\alpha/2}\,\stderr(\hat\theta)]$ where $t_{2,1-\alpha/2}$ is the $(1-\alpha/2)$-quantile of the Student-$t$ distribution with $2$~degrees of freedom, and divided into equally spaced grids. Once $\theta_0$ meets the condition, the procedure is stopped. At this grid, the optimal boundary has been identified as $\theta_0$. 
The procedure is similar for the lower bound, except that the direction of search starts from the largest grid to the smallest grid, and the test to consider in Step 3 is $H_{0}:\theta=\theta_0\ \text{vs.} \ H_{1}:\theta>\theta_0,$. In summary, see Table \ref{tab:algorithm2} for the pseudo code of the second proposed studentized permutation-based confidence interval. Therefore, the permutation $100\%(1-\alpha)$ confidence interval $\ciPtwo$ for $\theta$ is defined as
\begin{equation}\label{ci-p2}
    [\theta_{lower}; \theta_{upper}]\nonumber
\end{equation}

Since this approach uses a permutation test for every value of $\theta_0$, it is more computationally expensive than the previously mentioned method. Increasing the number of grids might reduce the length of confidence intervals. However, the wider the range for $\theta_0$ or the greater the number of grids, the higher the computational expense is. In practice, we suggest choosing $[\hat\theta+z_{1-\alpha/2}\stderr(\hat\theta);\hat\theta+t_{2,1-\alpha/2}\stderr(\hat\theta)]$ and $[\hat\theta-t_{2,1-\alpha/2}\stderr(\hat\theta);\hat\theta-z_{1-\alpha/2}\stderr(\hat\theta)]$ as the ranges for searching the upper bound and the lower bound, respectively. 

\begin{table}[]
    \centering
    \caption{Pseudo code for constructing the upper bound of the search- and permutation-based $100\%(1-\alpha)$ confidence interval. Construction of the lower bound works analogously.}
    \small
    \begin{tabular}{ll}
        \toprule
        \multicolumn{2}{l}{\texttt{For} $\theta_{0i}=i\times(b-a)/(M-1),i=1,\cdots,M$, \texttt{where $M$ denotes}}\\
        \multicolumn{2}{l}{\texttt{the number of grid points in} $[a;b],b\ge a\ge\hat\theta$,}\\
        1. & \texttt{Compute} $\bm{Y}_{\theta_{0i}}=\bm{Y}-\theta_{0i}\bm{Z}\bm{1}_k$.\\
        2. & \texttt{Replace} $\bm{Y}$ \texttt{with} $\bm{Y}_{\theta_{0i}}$ \texttt{and perform steps 1-3 in Table} \ref{tab:algorithm}.\\
        3. & \texttt{Compute p-value for the test $H_{0}:\theta=\theta_0\ \texttt{vs.} \ H_{1}:\theta<\theta_0,$:} \\
           & $\text{p-value}=1-\sum_{p=1}^{N}\mathbbm{1}(t_{(p)}\ge t)/N.$\\
         4.  & \texttt{If} $\text{p-value}\le\alpha/2$\texttt{, stop and return} $\theta_{upper}:=\theta_{0i}$.\\
         \bottomrule
    \end{tabular}
    \label{tab:algorithm2}
\end{table}

\section{Simulation}\label{sec:sim}
\subsection{Aims}
We investigate the small-sample properties of the proposed permutation technique in comparison with the competing procedures described in Section~\ref{sec:preliminary}.
Since we expect permutation approaches to be advantageous in particular in cases of few small studies, we consider the case of meta-analyses of $k=4$~studies and sample sizes in a range from~$n_i=15$ up to~$200$ participants.

\subsection{Settings}
Data are simulated from the model as specified in~(\ref{eqn:mod}) under different scenarios of varying parameters. The detailed parameter settings are shown in Table~\ref{tab:sim}. The number of studies is set to $4$, corresponding to a small meta-analysis. The size of studies is drawn uniformly from a range from $n_i=30$ up to~$100$ and from~$100$ to~$200$ to cover meta-analyses of \emph{small} and \emph{medium} studies, respectively. Moreover, the size of studies is set to follow a mixture of distribution $0.8 \unifdistn(15, 30) + 0.2 \unifdistn(30, 100)$ to consider meta-analyses of very small studies and guarantee the variation in study sizes.
The treatment effect is set to~$\theta=0$ or~$1$ to reflect the null and alternative hypotheses, respectively. The residual variances are set either to $\sigma_i=1$ for all included studies, or to differing values to account for cases of equal and unequal residual variances.
Moreover, the residual variances in all the cases are set such that the UISD of meta-analysis is equal to one.
Between-study heterogeneity is varied from $\tau=0.01$ (near homogeneity) up to $\tau=1.0$  
to cover up the range from small to large heterogeneity \cite{roever_et_al21}.
The proportion of the treated patients, $p_i$, is varied between~$0.5$ to~$0.7$, to include  balanced and unbalanced treatment allocation. 

Model~(\ref{eqn:mod}) is fitted to the data (using REML estimation, and assuming equal residual variances across studies).
The proposed permutation approaches from Section~\ref{sec:method} are applied, using $N=10\,000$ permutations for the test and intervals~$\ciPone$, and $2\,000$~iterations for the intervals~$\ciPtwo$
to reduce the computational expense in the simulation study, following \citet{lee_braun11}, who used $2\,000$~permutations for testing random effects in linear mixed models.


In our simulation study, the performance of the test is evaluated via its observed type~I error and power, while the performance of the confidence intervals is assessed via their empirical coverage probability and average length.
The number of grid points~$M$ in the computation of~$\ciPtwo$ (shown in Table~\ref{tab:algorithm2}) is set to~$5$,
starting from $\hat\theta+z_{(1-\alpha/2)}\,\stderr(\hat\theta)$
to $\hat\theta+4\,\stderr(\hat\theta)$ for the upper bound and 
from $\hat\theta-4\,\stderr(\hat\theta)$ 
to $\hat\theta-z_{(1-\alpha/2)}\,\stderr(\hat\theta)$ for the lower bound, respectively. 
To examine the robustness of the permutation test to the residual mis-specification, residuals are also set to have different variances or to be distributed as a Student-$t_{3}$ distribution or a $\mbox{log-Normal}(0, 1)$ distribution (to cover cases of heavy-tailed or skewed residuals). When residuals are not normally distributed, they are scaled such that their variances are equal to the variances from the normal case. The number of replicates is set to $1\,000$. The simulation is performed in R programming language \cite{r24} and the code is available in \url{https://github.com/davidtran2908/permu.git}.

In this section, we restrict our analysis to particular sample size involving normal error with equal variances, log-normal error, and Student-$t$ distributed error. More detailed simulation results for very small and medium study size, and normal error with unequal variances can be found in the supplementary material.

\begin{table}[t]
\centering
\caption{Parameter settings for the simulations discussed in Section~\ref{sec:sim}.}\label{tab:sim}
\small
\begin{tabular}{rcl}
  \toprule
  \multicolumn{2}{r}{parameter}        & settings \\
  \midrule
  number of studies    & $k$           & {$4$}\\
  study size           & $n_{i}$       & medium: $\unifdistn(100, 200)$,\\
                       &               & small: $\unifdistn(30, 100)$, \\
                       &               & very small: $0.8 \unifdistn(15, 30)+0.2 \unifdistn(30, 100)$ \\
  treatment allocation & $p_{i}$       & {$\unifdistn(0.5, 0.7)$}\\
  intercepts           & $\beta_{0i}$  & $(0.9, 2.3, 0.3, 0.1)^\top$\\
  slopes               & $\beta_{1i}$  & $(0.8, 0.7, 0.9, 0.9)^\top$ \\
  treatment effect     & $\theta$      & $\{0, 1\}$\\
  heterogeneity        & $\tau$        & $\{0.01, 0.1, 0.3, 0.5, 0.7, 1.0\}$\\
  residual s.d.    & $\sigma_{i}$  & $1, (0.9, 0.9, 0.9, 1.4)^\top$\\
  baseline value       & $y_{0ij}$     & $\normaldistn(4, 1)$\\
  residuals            & $\epsilon_{ij}$ & $\normaldistn$, Student-$t_3$, log-Normal \\
  permutations         & $N$           & $10\,000$ (permutation test, $\ciPone$), \\
                       &               & $2\,000$ (2nd permutation, $\ciPtwo$)\\
  \bottomrule
\end{tabular}
\end{table}

\subsection{Results}

In the following, we present the simulation results for different error distributions, including Normal, Student-$t_{3}$ and log-Normal residuals, under small sample sizes in each study. Additional results, including those for very small sample size, medium sample size, and unequal residual variances scenarios, are provided in the supplementary material.

\begin{figure}[hpt!]
  \centering
  \includegraphics[width=380pt]{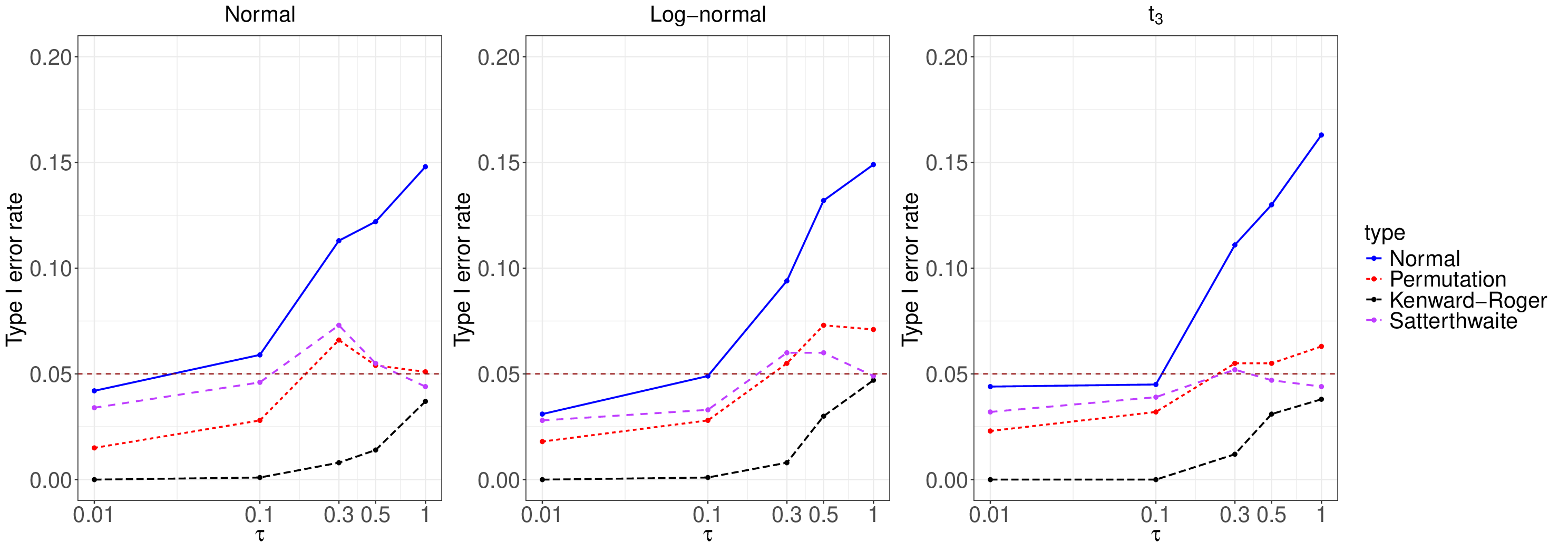}\\
  \caption{Simulated type-I error of the discussed analysis methods (colour coded), for various scenarios; Normal, log-Normal and Student-$t$ residuals (3~panels), between-study heterogeneity~$\tau$ ranged from~$0.01$ to~$1.0$ ($x$-axis).
  Studies were of ``small'' sample size, the nominal level (5\%) is shown by the dotted straight line.}
  \label{type-i-error-rate}
\end{figure}

\begin{figure}[hpt!]
  \centering
  \includegraphics[width=380pt]{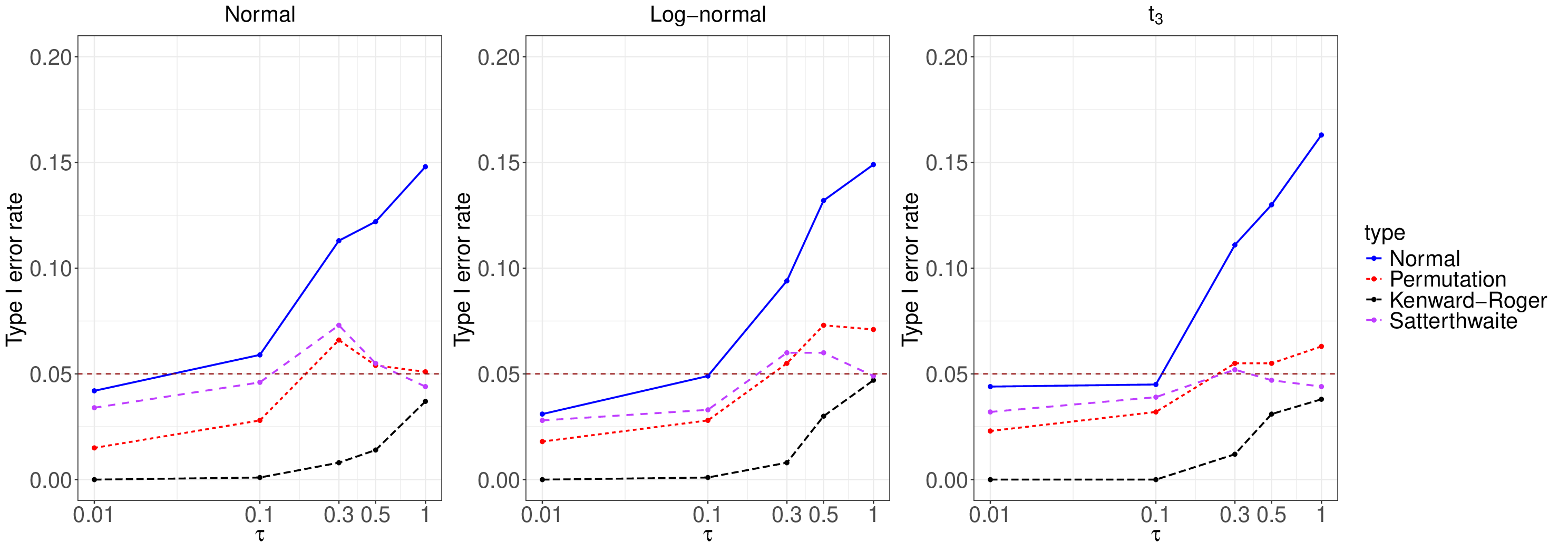}\\
  \caption{Power ($y$-axis) of the discussed analysis methods (colour-coded) for various scenarios;
  Normal, log-Normal and Student-$t$ residuals (3~panels) between-study heterogeneity~$\tau$ ranged from~$0.01$ to~$1.0$ ($x$-axis).
  Studies were of ``small'' sample size, and the nominal type-I error is at~5\%.}
  \label{power}
\end{figure}

We first examine the type-I error and power across different methods. The simulation results are illustrated in Figures~\ref{type-i-error-rate} and~\ref{power}.
From Figure~\ref{type-i-error-rate}, we find that Kenward-Roger’s approach tend to be conservative and the Normal approximation tends to be liberal, when $\tau$~varies from~$0.01$ up to~$1$. When $\tau$~is small, all methods are conservative. For most $\tau$~values, the Normal approximation as well as Kenward-Roger’s approach fail to control the type-I error.  In addition, we observe that the Normal approach slightly outperforms other methods in controlling the type-I error when $\tau$~is less than 0.1, \emph{and} the error follows normal distribution. However, the Normal approach fails to control type-I error and yields slightly more liberal results when $\tau>0.3$. This indicates its limitations in scenarios with high between-study heterogeneity~$\tau$. When $\tau$~increases, the proposed permutation method controls the type-I error better than other methods. On average, the proposed method controls the type-I error across a broader range of heterogeneity levels.


From Figure \ref{power}, we find that the proposed permutation method has a higher power when the normality assumption is violated, as in the log-Normal scenario. Although the type-I error of our proposed method is similar to that of Satterthwaite’s method, the power of our proposed method under certain alternatives outperforms that of Satterthwaite’s method in some simulation scenarios. The Normal approximation appears to provide the best power, however, since the type-I error it not controlled, these figures are not really comparable.

\begin{figure}[hpt!]
  \centering
  \includegraphics[width=380pt]{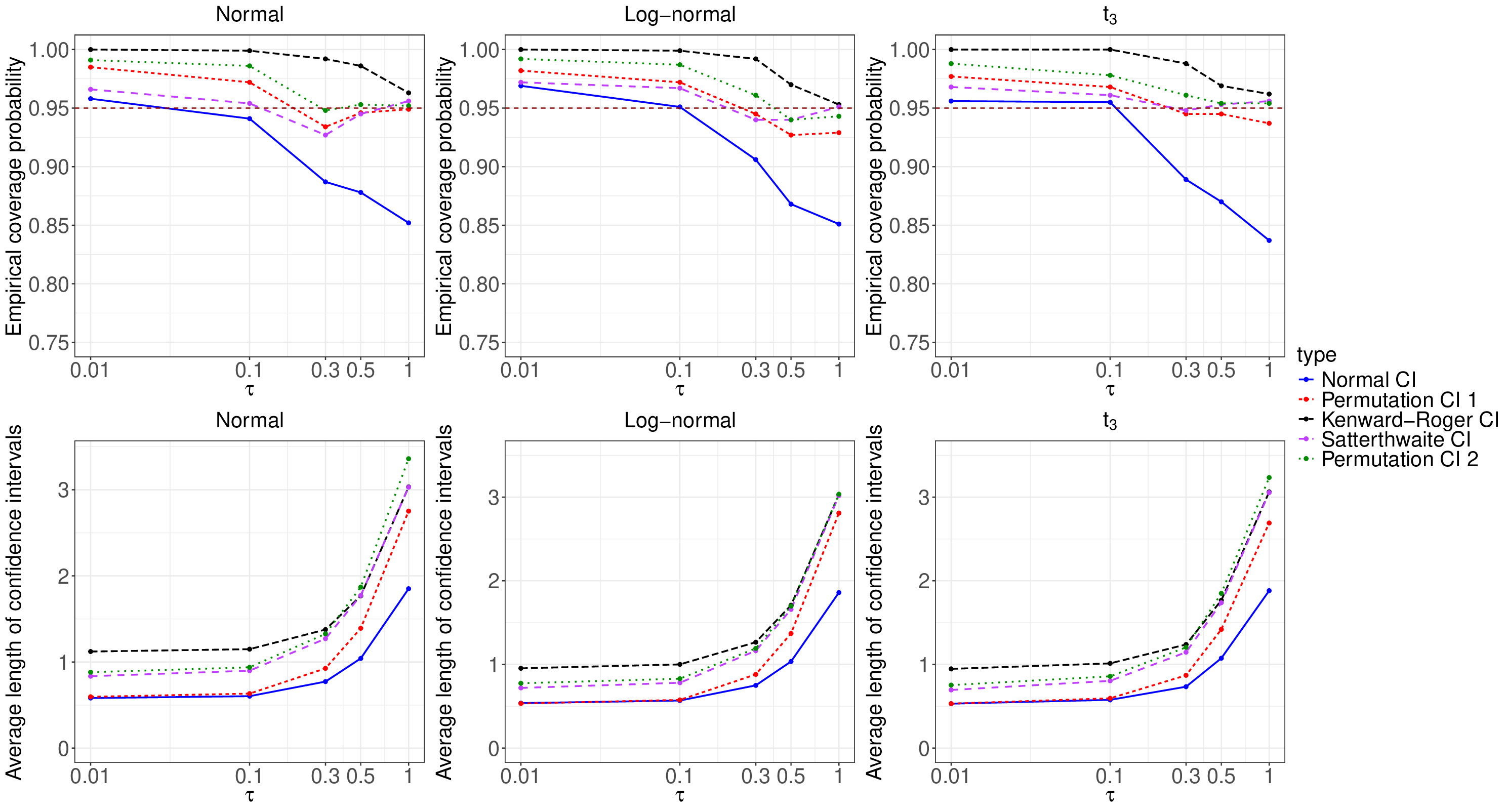}\\
  \caption{Simulated empirical coverage probability (top) and average length of confidence intervals (bottom) of the different analysis methods for various scenarios under the Null hypothesis ($\theta=0$); residuals are Normal, log-Normal or Student-$t_3$ distributed (left to right) and between-study heterogeneity~$\tau$ ranges up to~$1.0$ ($x$-axis). 
  All studies are of ``small'' size, the nominal level of~$5\%$ is indicated by the dotted horizontal line.}
  \label{ecp-al-theta0}
\end{figure}

\begin{figure}[hpt!]
  \centering
  \includegraphics[width=380pt]{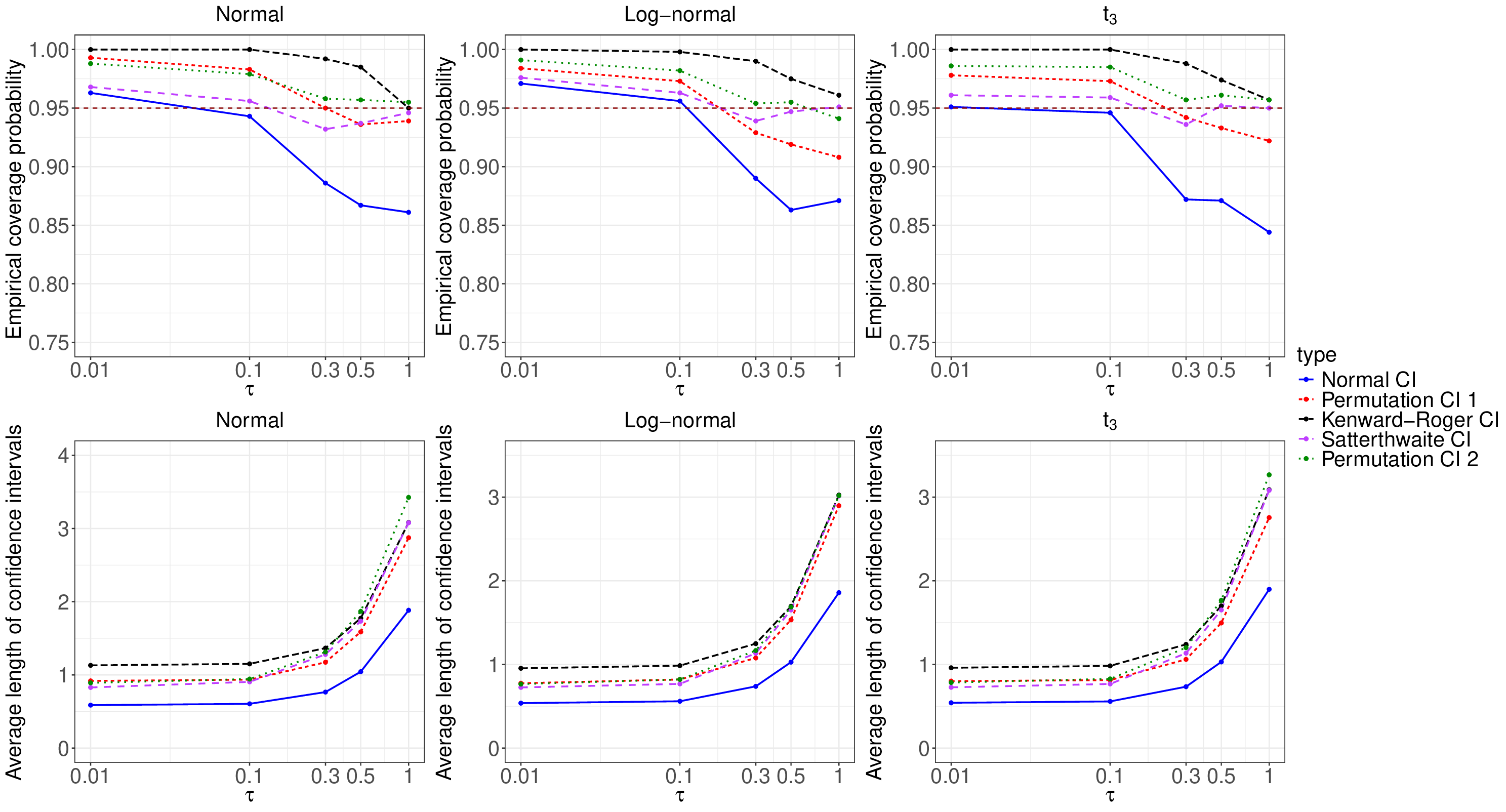}\\
  \caption{Simulated empirical coverage probability (top) and average length of confidence intervals (bottom) of the different analysis methods for various scenarios under the alternative hypothesis ($\theta=1$); residuals are Normal, log-Normal or Student-$t_3$ distributed (left to right) and between-study heterogeneity~$\tau$ ranges up to~$1.0$ ($x$-axis). 
  All studies are of ``small'' size, the nominal level of~$5\%$ is indicated by the dotted horizontal line.}
  \label{ecp-al-theta1}
\end{figure}

Next, we compare confidence intervals across these methods. For the permutation test, we propose two different approaches to construct confidence intervals. One ($\ciPone$) is based on the percentiles of the permutation distribution of the effect estimator, while the other ($\ciPtwo$) uses a search algorithm to find the optimal confidence interval boundaries that maintain the nominal level. We first examine the empirical coverage probability and average length of confidence intervals across different methods. The simulation results are illustrated in Figure~\ref{ecp-al-theta0} under~$\theta=0$ and Figure~\ref{ecp-al-theta1} under~$\theta=1$.

From Figure \ref{ecp-al-theta0} and Figure \ref{ecp-al-theta1}, we find that the permutation methods perform better than the existing methods in terms of empirical coverage probability. While the permutation confidence intervals $\ciPone$ and $\ciPtwo$ maintain a similar empirical coverage probability compared to Satterthwaite’s method, their average lengths are smaller. In contrast, the Normal confidence interval shows a much lower empirical coverage probability as $\tau$ increases. Overall, the permutation confidence interval $\ciPone$ maintains the empirical coverage probability while achieving a shorter average length. 
Due to the boundary-searching procedure, the average length of the permutation confidence interval $\ciPtwo$ tends to be longer than that of $\ciPone$.
In all scenarios, the average length of the confidence interval constructed by all methods is increasing with the increasing of $\tau$. It indicates that larger heterogeneity between studies leads to increased uncertainty and results in wider confidence intervals.
In addition, the Kenward-Roger confidence interval ($\ciKR$) has the highest empirical coverage probability, which deviates from the nominal level, and also has the largest average confidence interval length in almost all scenarios.

For normal outcomes, the permutation confidence interval $\ciPtwo$ performs much better than the permutation confidence interval $\ciPone$ in terms of empirical coverage probability. However, the permutation confidence interval $\ciPtwo$ tends to have a larger average confidence interval length compared to the permutation confidence interval $\ciPone$ as $\tau$~increases.






\section{Discussion}\label{sec:discussion}

In this paper, we propose a studentized permutation test for the treatment effect in IPD meta-analysis of continuous outcomes.
Use of a permutation approach is facilitated by enforcing exchangeability of the standardized residuals (despite potentially varying residual variances among included trials).
The proposed method does not rely on normality of outcomes or large-sample asymptotics, and works well for small meta-analyses (few small studies).
Based on this test, two types of confidence intervals are derived. One confidence interval ($\ciPone$) is computed based on the percentiles of the permutation distribution of the effect estimator, while the second type ($\ciPtwo$) utilizes a search across parameter space to identify values closest to the treatment effect estimate such that they are just significantly different from the true treatment effect. Notably, the second type does not violate the exchangeability assumption, even when the true effect is substantially different from zero, because standardized residuals are permuted \emph{after} centering the outcome vector. The simulation results show that the proposed confidence intervals exhibit good performance in a range of scenarios.

We provide simulation results to support the validity of the proposed method and demonstrate that it outperforms existing approaches. 
Moreover, the computational expense of the search-based confidence interval method should be reduced in the future, for example, by improving on the search algorithm.

It should be straightforward to extend the proposed method to cases when there is more than one random effect, e.g., when the intercept in model~(\ref{eqn:mod}) is random instead of stratified by study \cite{riley_debray21}, as long as the exchangeability assumption of the standardized residuals holds. Another direction for future work, which is more challenging, is to extend the proposed method to other types of outcomes in IPD meta-analysis, as such data are typically modeled using generalized linear mixed models (GLMMs). 



\section*{Acknowledgments}
Support from the \emph{Deutsche Forschungsgemeinschaft (DFG)} is gratefully acknowledged (grant number \mbox{FR 3070/3-2}).

%





\bibliography{refs}

\begin{thebibliography}{26}
\providecommand{\natexlab}[1]{#1}
\providecommand{\url}[1]{\texttt{#1}}
\expandafter\ifx\csname urlstyle\endcsname\relax
  \providecommand{\doi}[1]{doi: #1}\else
  \providecommand{\doi}{doi: \begingroup \urlstyle{rm}\Url}\fi

\bibitem[Basso and Finos(2012)]{basso_finos12}
D.~Basso and L.~Finos.
\newblock Exact multivariate permutation tests for fixed effects in
  mixed-models.
\newblock \emph{Communications in Statistics - Theory and Methods}, 41\penalty0
  (16-17):\penalty0 2991--3001, 2012.

\bibitem[Debray et~al.(2015)Debray, Moons, van Valkenhoef, Efthimiou, Hummel,
  Groenwold, Reitsma, and {on behalf of the GetReal methods review
  group}]{debray_et_al15}
T.~P.~A. Debray, K.~G.~M. Moons, G.~van Valkenhoef, O.~Efthimiou, N.~Hummel,
  R.~H.~H. Groenwold, J.~B. Reitsma, and {on behalf of the GetReal methods
  review group}.
\newblock Get real in individual participant data ({IPD}) meta-analysis: a
  review of the methodology.
\newblock \emph{Research Synthesis Methods}, 6\penalty0 (4):\penalty0 293--309,
  2015.

\bibitem[Follmann and Proschan(2004)]{follmann_proschan04}
D.~A. Follmann and M.~A. Proschan.
\newblock Valid inference in random effects meta-analysis.
\newblock \emph{Biometrics}, 55\penalty0 (3):\penalty0 732--737, 05 2004.

\bibitem[Giesbrecht and Burns(1985)]{giesbrecht_burns85}
F.~G. Giesbrecht and J.~C. Burns.
\newblock Two-stage analysis based on a mixed model: Large-sample asymptotic
  theory and small-sample simulation results.
\newblock \emph{Biometrics}, 41\penalty0 (2):\penalty0 477--486, 1985.

\bibitem[Goldstein et~al.(2000)Goldstein, Yang, Omar, Turner, and
  Thompson]{goldstein_et_al00}
H.~Goldstein, M.~Yang, R.~Omar, R.~Turner, and S.~Thompson.
\newblock Meta-.
\newblock \emph{Journal of the Royal Statistical Society. Series C (Applied
  Statistics)}, 49\penalty0 (3):\penalty0 399--412, 2000.

\bibitem[Halekoh and Højsgaard(2014)]{halekoh_højsgaard14}
U.~Halekoh and S.~Højsgaard.
\newblock A {K}enward-{R}oger approximation and parametric bootstrap methods
  for tests in linear mixed models – {T}he {R} package pbkrtest.
\newblock \emph{Journal of Statistical Software}, 59\penalty0 (9):\penalty0
  1–32, 2014.

\bibitem[Hedges and Olkin(1985)]{hedges_olkin85}
L.~V. Hedges and I.~Olkin.
\newblock Introduction.
\newblock In L.~V. Hedges and I.~Olkin, editors, \emph{Statistical Methods for
  Meta-Analysis}, chapter~1, pages 1--14. Academic Press, San Diego, 1985.

\bibitem[Higgins and Green(2011)]{higgins_green11}
J.~Higgins and S.~Green.
\newblock \emph{Cochrane Handbook for Systematic Reviews of Interventions}.
\newblock Cochrane Collaboration, 2011.

\bibitem[Kackar and Harville(1984)]{kackar_harville84}
R.~N. Kackar and D.~A. Harville.
\newblock Approximations for standard errors of estimators of fixed and random
  effect in mixed linear models.
\newblock \emph{Journal of the American Statistical Association}, 79\penalty0
  (388):\penalty0 853--862, 1984.

\bibitem[Kambič et~al.(2024)Kambič, Feuerstein, Tran, Friede, Edelmann,
  Lainscak, and {ExTraLVAD Collaborators}]{kambic_et_al24}
T.~Kambič, A.~Feuerstein, P.~T. Tran, T.~Friede, F.~Edelmann, M.~Lainscak, and
  {ExTraLVAD Collaborators}.
\newblock Exercise training in left ventricular assist device patients:
  Protocol of an individual participant data meta-analysis.
\newblock \emph{ESC Heart Failure}, 2024.

\bibitem[Kenward and Roger(1997)]{kenward_roger97}
M.~G. Kenward and J.~H. Roger.
\newblock Small sample inference for fixed effects from restricted maximum
  likelihood.
\newblock \emph{Biometrics}, 53\penalty0 (3):\penalty0 983--997, 1997.

\bibitem[Kherad-Pajouh and Renaud(2010)]{kheradpajouh_renaud10}
S.~Kherad-Pajouh and O.~Renaud.
\newblock An exact permutation method for testing any effect in balanced and
  unbalanced fixed effect anova.
\newblock \emph{Computational Statistics \& Data Analysis}, 54\penalty0
  (7):\penalty0 1881--1893, 2010.

\bibitem[Kuznetsova et~al.(2017)Kuznetsova, Brockhoff, and
  Christensen]{kuznetsova_et_al17}
A.~Kuznetsova, P.~B. Brockhoff, and R.~H.~B. Christensen.
\newblock lmertest package: Tests in linear mixed effects models.
\newblock \emph{Journal of Statistical Software}, 82\penalty0 (13):\penalty0
  1–26, 2017.

\bibitem[Lee and Braun(2011)]{lee_braun11}
O.~E. Lee and T.~M. Braun.
\newblock Permutation tests for random effects in linear mixed models.
\newblock \emph{Biometrics}, 68\penalty0 (2):\penalty0 486--493, 2011.

\bibitem[Legha et~al.(2018)Legha, Riley, Ensor, Snell, Morris, and
  Burke]{legha_et_al18}
A.~Legha, R.~D. Riley, J.~Ensor, K.~I.~E. Snell, T.~P. Morris, and D.~L. Burke.
\newblock Individual participant data meta-analysis of continuous outcomes: {A}
  comparison of approaches for specifying and estimating one-stage models.
\newblock \emph{Statistics in Medicine}, 37\penalty0 (29):\penalty0 4404--4420,
  2018.

\bibitem[Nyblom(2015)]{nyblom15}
J.~Nyblom.
\newblock Permutation tests in linear regression.
\newblock In K.~Nordhausen and S.~Taskinen, editors, \emph{Modern
  Nonparametric, Robust and Multivariate Methods: Festschrift in Honour of
  Hannu Oja}, pages 69--90. Springer International Publishing, Cham, 2015.

\bibitem[O'Gorman(2001)]{ogorman01}
T.~W. O'Gorman.
\newblock An adaptive permutation test procedure for several common tests of
  significance.
\newblock \emph{Computational Statistics \& Data Analysis}, 35\penalty0
  (3):\penalty0 335--350, 2001.

\bibitem[Pinheiro(2000)]{pinheiro00}
J.~C. Pinheiro.
\newblock Theory and computational methods for linear mixed-effects models.
\newblock In \emph{Mixed-Effects Models in S and S-PLUS}, pages 57--96.
  Springer New York, New York, NY, 2000.

\bibitem[{R Core Team}(2024)]{r24}
{R Core Team}.
\newblock \emph{R: A Language and Environment for Statistical Computing}.
\newblock R Foundation for Statistical Computing, Vienna, Austria, 2024.

\bibitem[Riley and Debray(2021)]{riley_debray21}
R.~D. Riley and T.~P.~A. Debray.
\newblock The one-stage approach to ipd meta-analysis.
\newblock In \emph{Individual Participant Data Meta‐Analysis}, chapter~6,
  pages 127--162. John Wiley \& Sons, Ltd, 2021.

\bibitem[Riley et~al.(2013)Riley, Kauser, Bland, Thijs, Staessen, Wang,
  Gueyffier, and Deeks]{riley_et_al13}
R.~D. Riley, I.~Kauser, M.~Bland, L.~Thijs, J.~A. Staessen, J.~Wang,
  F.~Gueyffier, and J.~J. Deeks.
\newblock Meta-analysis of randomised trials with a continuous outcome
  according to baseline imbalance and availability of individual participant
  data.
\newblock \emph{Statistics in Medicine}, 32\penalty0 (16):\penalty0 2747--2766,
  2013.

\bibitem[Riley et~al.(2021)Riley, Tierney, and Stewart]{RileyTierneyStewart}
R.~D. Riley, J.~F. Tierney, and L.~A. Stewart, editors.
\newblock \emph{Individual participant data meta-analysis: A handbook for
  healthcare research}.
\newblock Wiley, Hoboken, NJ, USA, 2021.

\bibitem[Röver et~al.(2021)Röver, Bender, Dias, Schmid, Schmidli, Sturtz,
  Weber, and Friede]{roever_et_al21}
C.~Röver, R.~Bender, S.~Dias, C.~H. Schmid, H.~Schmidli, S.~Sturtz, S.~Weber,
  and T.~Friede.
\newblock On weakly informative prior distributions for the heterogeneity
  parameter in {B}ayesian random-effects meta-analysis.
\newblock \emph{Research Synthesis Methods}, 12\penalty0 (4):\penalty0
  448--474, 2021.

\bibitem[Satterthwaite(1946)]{satterthwaite46}
F.~E. Satterthwaite.
\newblock Approximate distribution of estimates of variance components.
\newblock \emph{Biometrics Bulletin}, 2\penalty0 (6):\penalty0 110--114, 1946.

\bibitem[Schmid et~al.(2020)Schmid, Stijnen, and White]{schmid_et_al20}
C.~Schmid, T.~Stijnen, and I.~White.
\newblock \emph{Handbook of Meta-Analysis}.
\newblock Chapman and Hall/CRC, 1st edition, 2020.

\bibitem[Winkler et~al.(2014)Winkler, Ridgway, Webster, Smith, and
  Nichols]{winkler_et_al14}
A.~M. Winkler, G.~R. Ridgway, M.~A. Webster, S.~M. Smith, and T.~E. Nichols.
\newblock Permutation inference for the general linear model.
\newblock \emph{NeuroImage}, 92:\penalty0 381--397, 2014.

\end{thebibliography}

\newpage

\section*{Appendix}

\begin{figure}[hpt!]
  \centering
  \includegraphics[width=380pt]{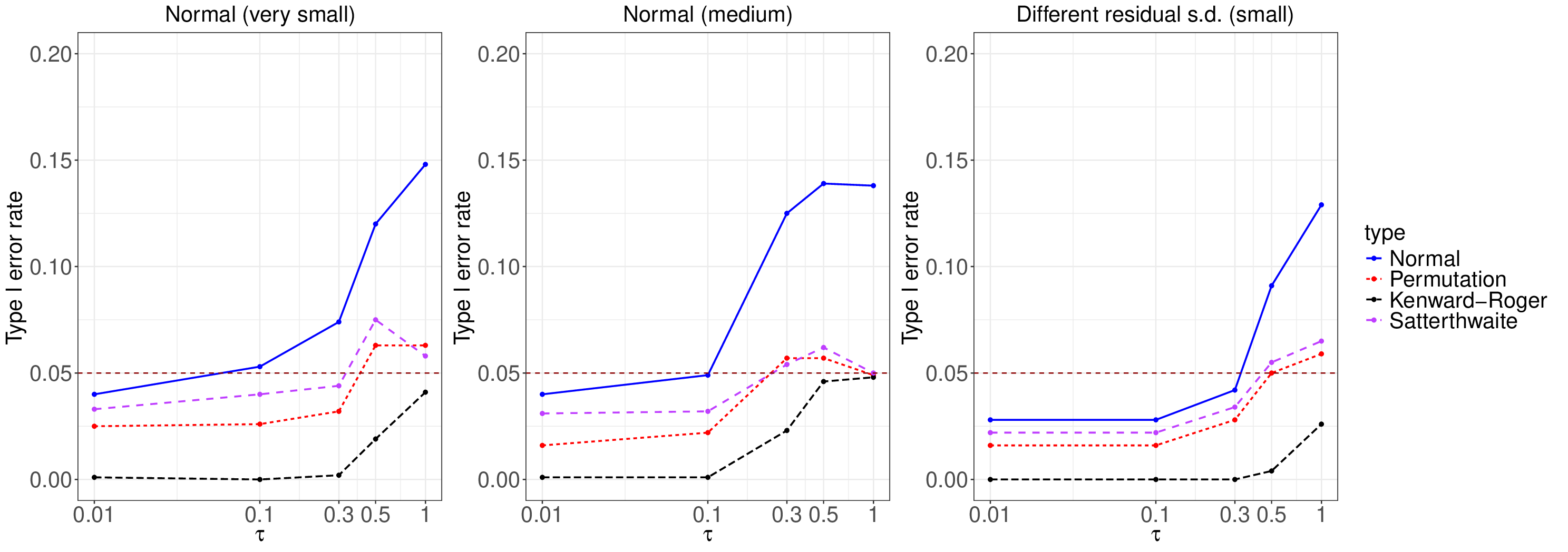}\\
  \caption{Simulated type-I error of the discussed analysis methods (colour coded), for various scenarios; ``very small'' sample size, ``medium'' sample size, varying residual variances, between-study heterogeneity~$\tau$ ranged from~$0.01$ to~$1.0$ ($x$-axis).
  Residuals are Normally distributed, the nominal level (5\%) is shown by the dotted straight line.}
  \label{type-i-error-rate-2}
\end{figure}

\begin{figure}[hpt!]
  \centering
  \includegraphics[width=380pt]{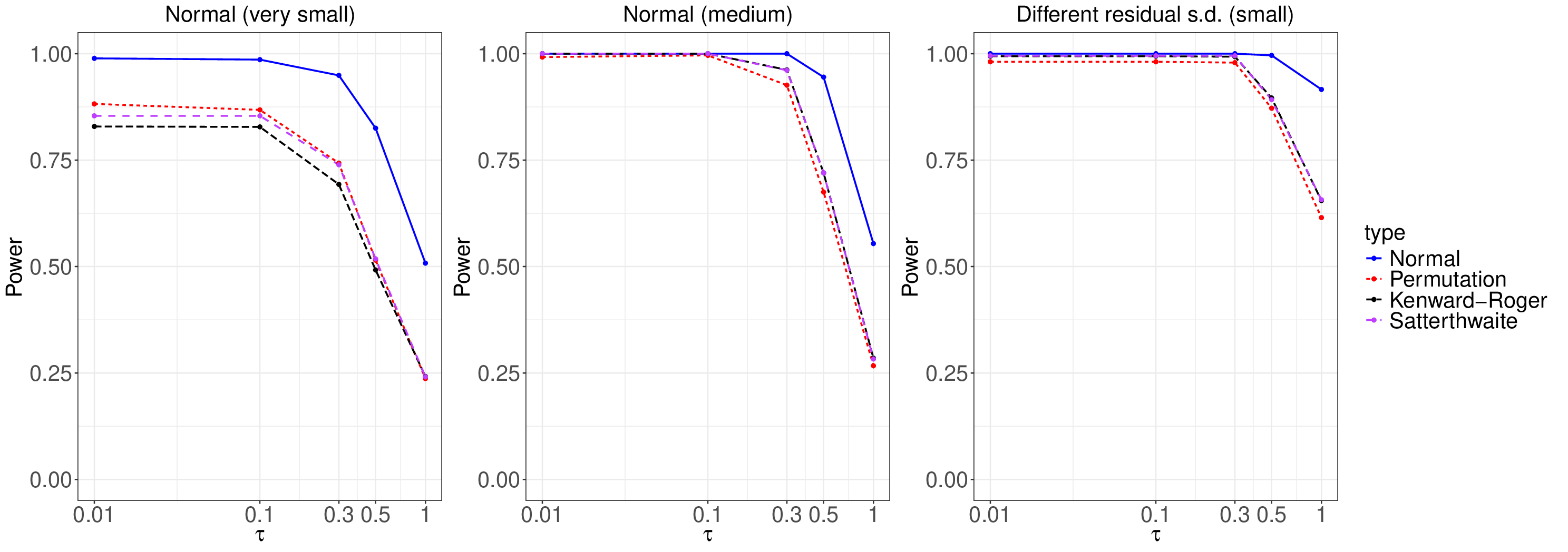}\\
  \caption{Power ($y$-axis) of the discussed analysis methods (colour-coded) for various scenarios;
  ``small'' sample size, ``medium'' sample size, varying residual variances (3~panels) between-study heterogeneity~$\tau$ ranged from~$0.01$ to~$1.0$ ($x$-axis).
  Residuals are Normally distributed, and the nominal type-I error is at~5\%.
}
  \label{power-2}
\end{figure}

\begin{figure}[hpt!]
  \centering
  \includegraphics[width=380pt]{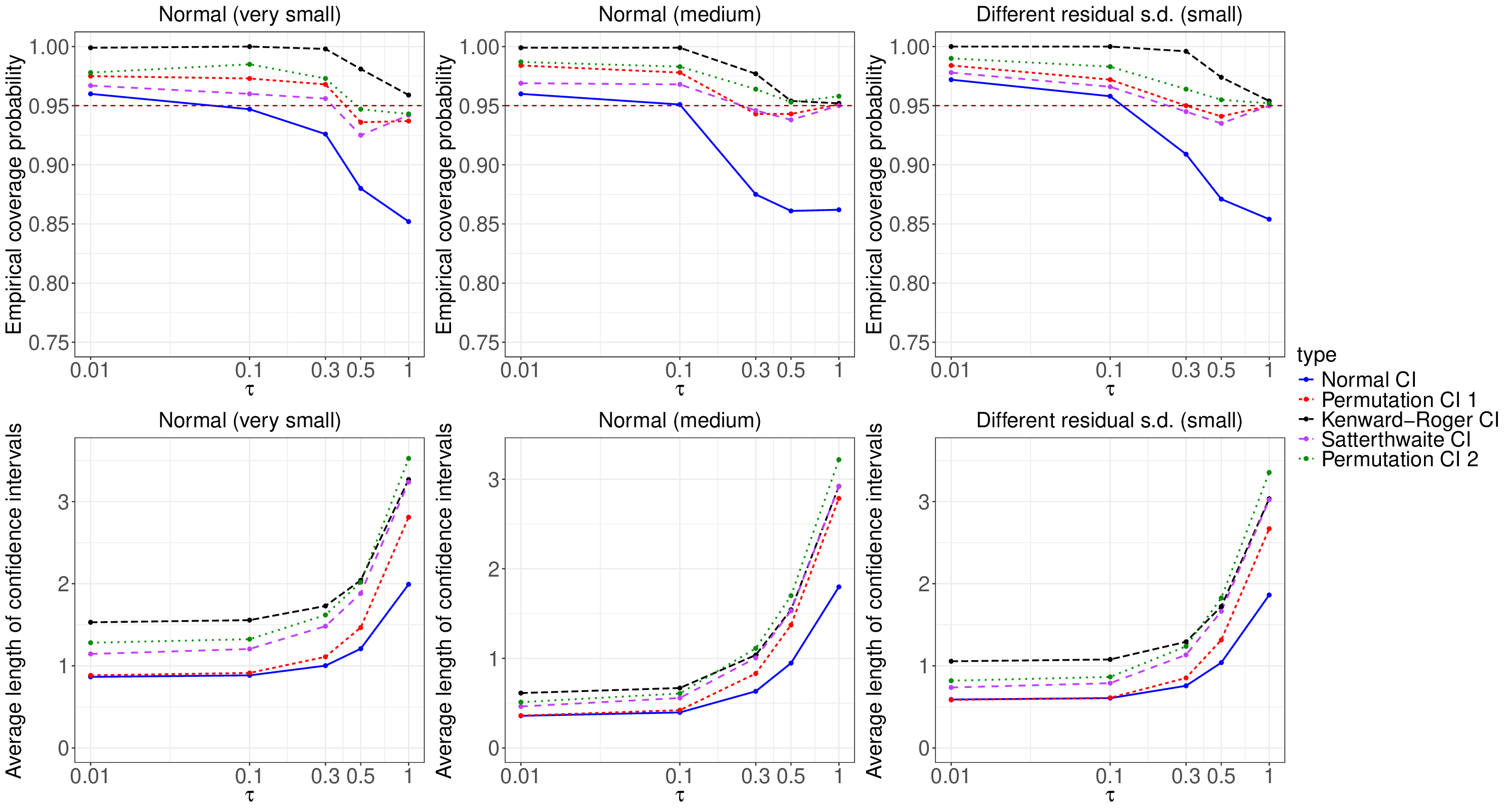}\\
  \caption{Simulated empirical coverage probability (top) and average length of confidence intervals (bottom) of the different analysis methods for various scenarios under the Null hypothesis ($\theta=0$); for various scenarios; ``small'' sample size, ``medium'' sample size, varying residual variances (left to right) and between-study heterogeneity~$\tau$ ranges up to~$1.0$ ($x$-axis). 
  Residuals are Normally distributed, the nominal level of~$5\%$ is indicated by the dotted horizontal line.
}
  \label{ecp-al-theta0_2}
\end{figure}

\begin{figure}[hpt!]
  \centering
  \includegraphics[width=380pt]{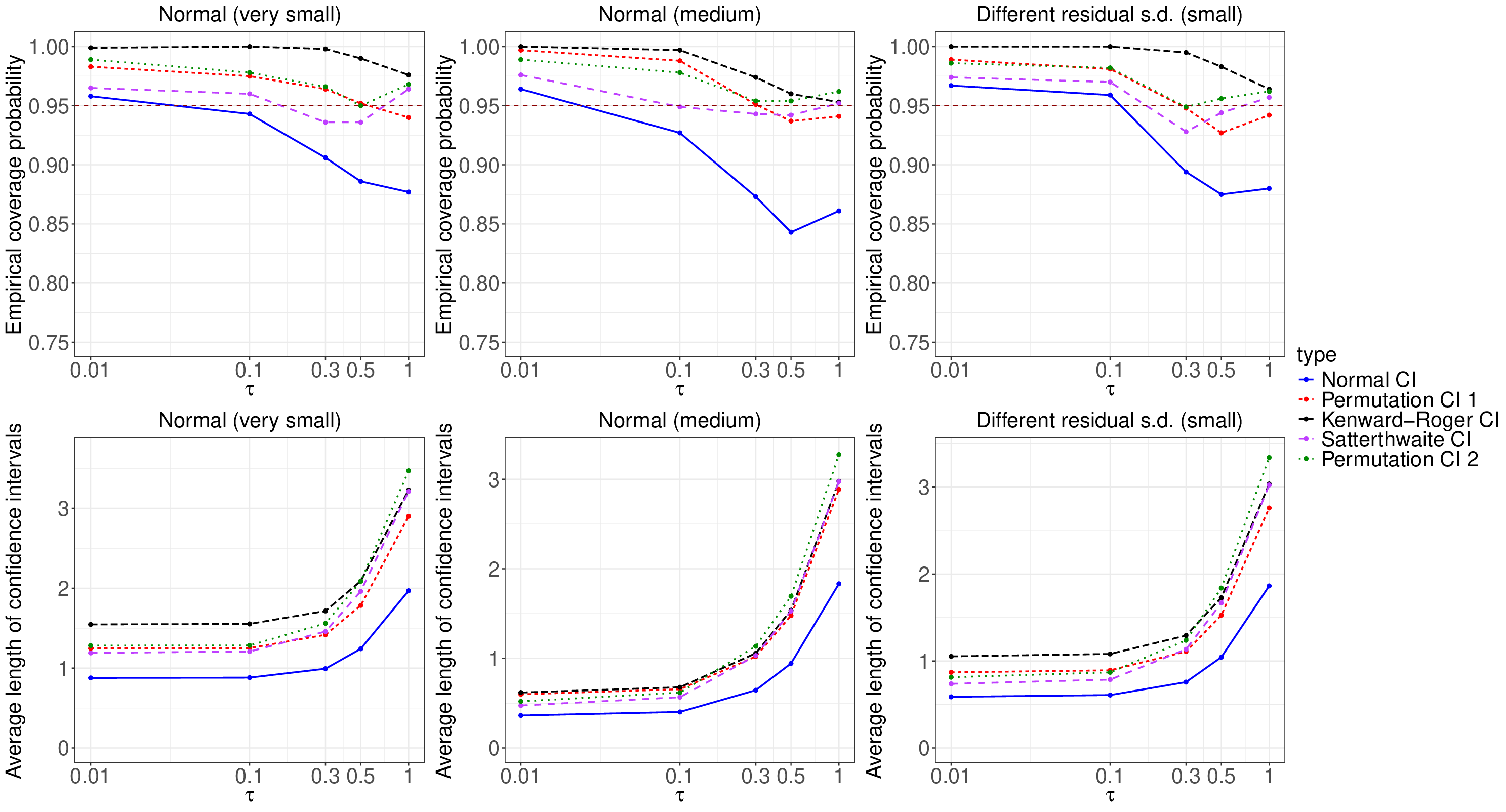}\\
  \caption{Simulated empirical coverage probability (top) and average length of confidence intervals (bottom) of the different analysis methods for various scenarios under the Null hypothesis ($\theta=1$); for various scenarios; ``small'' sample size, ``medium'' sample size, varying residual variances (left to right) and between-study heterogeneity~$\tau$ ranges up to~$1.0$ ($x$-axis). 
  Residuals are Normally distributed, the nominal level of~$5\%$ is indicated by the dotted horizontal line.}
  \label{ecp-al-theta1_2}
\end{figure}

Here we consider scenarios with normal errors under very small and medium sample sizes, as well as unequal residual variances. The type~I error and power are presented in Figure \ref{type-i-error-rate-2} and Figure \ref{power-2}, respectively. The empirical coverage probability and average length of the confidence interval are shown in Figure \ref{ecp-al-theta0_2} for a treatment effect of 0, and in Figure \ref{ecp-al-theta1_2} for a treatment effect of 1.

\end{document}